\documentclass[twocolumn,showpacs,preprintnumbers,amsmath,amssymb]{revtex4}
\usepackage{bm}
\usepackage{natbib}
\usepackage{graphicx}
\usepackage{dcolumn}
\usepackage{color} 
\begin{document}
\title{X-ray Dichroism in Noncentrosymmetryc crystals}
\author{Paolo Carra}
\email{carra@esrf.fr} 
\author{Andr\'{e}s Jerez} 
\author{Ivan Marri}
\altaffiliation[Also at ]{INFM and Dipartimento di Fisica 
 dell'Universit\`{a} degli Studi 
 di Modena e Reggio Emilia, 41100 Modena, Italy.} 
\affiliation{European Synchrotron Radiation Facility,
B.P. 220, F-38043 Grenoble C\'{e}dex, France}
\date{\today} 
\begin{abstract}
In this paper the authors analyse near-edge absorption of x rays 
in noncentrosymmetric crystals. The work is motivated by recent
observations of x-ray dichroic effects which stems from 
parity-nonconserving electron interactions. We provide a theoretical
description of these experiments and show that they are sensitive
to microscopic  {\it polar} and {\it magnetoelectric} properties of the 
sample. Our derivation extends previous theoretical work on 
centrosymmetric systems and identifies new directions in the
microscopic analysis of crystalline materials using x-ray absorption
spectroscopy.
\end{abstract}
\pacs{PACS numbers:78.70.Dm, 33.55.Ad}
\maketitle 
\section{Introduction}
\subsection{Pure electric transitions}
Near-edge x-ray dichroism with synchrotron radiation is a powerful 
probe of electronic states in crystals. As is known, its effectiveness 
stems from two prominent features of inner-shell excitations
\begin{itemize}
\item[] {\em site selectivity}, resulting from the tuning of the x-ray 
 energy at a given inner-shell threshold; 
\item[] electron {\em angular momentum resolution}, as enforced by the 
 selection rules
 of pure electric multipole (E1, E=2, ...) transitions, which raise 
 an inner-shell electron to empty valence orbitals.
\end{itemize}
Linear or circular polarisations are employed in experiments, 
leading to linear or circular x-ray dichroism, respectively. 
The former implies a difference between radiations with linear 
polarisations parallel or perpendicular to a local symmetry axis and is 
sensitive to charge anisotropies.
The latter measures the difference in absorption between right and
left circularly polarised radiations and reflects magnetic 
properties of crystals. 

Following the pioneering work of Templeton and Templeton \cite{TT80} 
and Sch\"{u}tz and coworkers \cite{Sch87}, a number of authors have 
recorded x-ray dichroic signals in a variety of samples, ranging across 
the periodic table from 3d transition metals to actinides. 

In parallel with this experimental activity, theory has aimed 
at identifying the microscopic origin of the observed spectra.
Working within an atomic model \cite{Tho92,Car93a,Car93b} (a good 
approximation, as it was later demonstrated \cite{BCA00}) a set of sum 
rules was obtained, which relate integrated dichroic intensities to the 
ground-state expectation value of effective {\it one-electron 
operators} (irreducible tensors) \cite{Note1}. Two classes of operators 
are obtained, which are identified 
by their transformation properties under space inversion (${\bf x} 
\rightarrow -{\bf x}$, $I$-transformation) and time reversal ($t\rightarrow -t$, 
$R$-transformation). They correspond to charge ($I$-even, $R$-even) 
and magnetic ($I$-even, $R$-odd) {\em order parameters} of crystals. X-ray 
dichroism is thus sensitive to long-range crystalline orderings; and these are 
distinguished by photon polarisation and by the nature (E1 or E2) of
the inner-shell excitation. Undoubtedly, the most important result
of this theoretical analysis has been to show that, in a ferromagnet (or 
ferrimagnet), x-ray circular dichroism provides a direct and independent 
determination of orbital and spin contributions to the magnetic moment 
\cite{Tho92,Car93a}. 

It is important to observe that pure electric multipole transitions cannot 
probe electronic properties which stem from the breaking of space inversion: 
all order parameters revealed by pure E1 and E2 transitions are 
even under the transformation $I$. 
\subsection{Interference}
In the x-ray region, other classes of dichroic phenomena have recently 
been investigated 
by Goulon {\it et al.} who reported the observation of three effects:
\begin{itemize}
\item[] {\it X-ray natural circular dichroism} (XNCD),
probed in 
Na$_3$Nd(digly)$_3\cdot$2NaBF$_4\cdot$6H$_2$O \cite{Ala98} and in
$\alpha$-LiIO$_3$ \cite{Gou98}. 
(The effect was observed near the Nd $L_3$ edge
and near the iodine $L$ edges.)
\item[]  {\it X-ray nonreciprocal linear dichroism} (XNLD),
detected near the vanadium $K$ edge in the low-temperature 
insulating phase of a Cr-doped V$_{2}$O$_{3}$ crystal\cite{Gou00}.
\item[] {\it X-ray magnetochiral dichroism} (XM$\chi$D), observed 
at the cromium K edge \cite{Gou02} in crystalline Cr$_2$O$_3$.
\end{itemize}
Here, inner-shell excitations are ascribed to the
E1-E2 interference; detecting a nonvanishing signal thus requires an ordered
structure and the breaking of space inversion.

In our view, the work of Goulon and his collaborators is of particular importance
as it identifies new directions in the microscopic analysis of materials
using x-ray absorption spectroscopy. In fact, symmetry considerations
indicate that XNCD, XNLD and XM$\chi$D are sensitive to {\it polar} and 
{\it magnetoelectric} (ME) properties of crystals.
\begin{table*}
\caption{\label{TableI} Order parameters and x-ray dichroic effects.}
\begin{ruledtabular}
\begin{tabular}{ccccc} 
Order Parameter  & Space Inversion & Time Reversal & 
X-ray Dichroism & X-ray Transition 
\\ \hline
Charge & even & even & Linear & Pure E1 or E2 \\ \hline 
Magnetic & even & odd & Magnetic Circular & Pure E1 or E2 \\ \hline
Electric & odd & even & Natural Circular & Interference E1-E2\\ 
\hline
Magnetoelectric & odd & odd & Nonreciprocal & Interference E1-E2\\ 
\end{tabular}
\end{ruledtabular}
\end{table*}

It is immediately obvious that the microscopic theory of integrated spectra
previously mentioned does not apply to the E1-E2 interference. Broader
symmetry considerations are required to identify order parameters 
probed by x-ray natural and
non-reciprocal dichroism. Such an analysis is outlined in 
the current work. Our discussion will be restricted to 
spectra integrated over the energy range corresponding to the two partners 
of a spin-orbit split inner shell: $j_{\pm} = l_c \pm \frac{1}{2}$, where 
$l_c$ denotes the angular momentum of the inner electron. Only 
orbital degrees of freedom (no spin) are observed in this case. 

It will be shown that the E1-E2 interference is described by $I$-odd 
microscopic operators, revealing the presence of parity nonconserving 
interactions. (We remind the reader that the expectation value 
of an $I$-odd operator vanishes in any state of definite parity.) As in 
the case of pure electric multipole transitions, such operators are further 
distinguished by their behaviour under time reversal. Two classes of 
order parameters are therefore identified
\begin{itemize}
\item[]
$R$-even and $I$-odd operators corresponding to one-electron  
polar properties, which arise from a 
noncentrosymmetric distribution of charge.
\item[]
$I$-odd and $R$-odd 
operators, obviously  invariant under the combined 
symmetry $RI$ and thus describing one-electron ME 
properties of crystals \cite{Dzy60}. 
\end{itemize}
It is thus readily seen that E1-E2 x-ray dichroism is sensitive to 
additional charge and magnetic orderings, which manifest themselves 
when the crystal lacks space-inversion symmetry. 
 
For convenience of the reader, a classification of order parameters 
as probed by various x-ray dichroisms is provided in Table 
\ref{TableI}. [Notice that a class may contain more than one 
 order parameter, as illustrated by the following example. Consider 
 pure E2 absorption. Four irreducible tensors of rank $k=0,1, ..., 4$ 
 contribute to the integrated spectrum, for arbitrary polarisation 
 \cite{Car93b}. In this case, magnetic circular dichroism selects 
 two order parameters: a vector ($k=1$) and an octapole ($k=3$).]
\section{E1-E2 x-ray dichroism}
This Section discusses general features of E1-E2 x-ray absorption. 
As we have anticipated, we will consider integrated spectra and express 
them as a linear combination of ground-state expectation values of 
one-electron irreducible tensors (order parameters). These order 
parameters are obtained by implementing Racah-Wigner recoupling 
techniques \cite{VMK88}, a convenient framework for dealing with 
the angular part of matrix elements. The ensuing results will then 
be applied to x-ray natural and nonreciprocal dichroism. 
Our discussion focusses on the physical interpetation of 
dichroic spectra; algebraic details of the derivation will
be given in an appendix.

\subsection{\label{ord-par} Space-odd order parameters}
Central to our considerations is the absorption cross section
\begin{widetext}
\begin{equation}
\sigma^{\bm \epsilon}(\omega) = 4\pi^2\alpha\hbar\omega
\left[
\frac{i}{2}\sum_f 
\sum_{\iota \iota'}
\langle g\mid {\bm \epsilon}^{\ast}\cdot {\bf r}_{\iota}
\mid f\rangle
\langle f\mid {\bm \epsilon} \cdot {\bf r}_{\iota'}
{\bf k}\cdot {\bf r}_{\iota'} \mid g\rangle 
+ {\rm c.c.}
\right]
\delta(E_f-E_g-\hbar\omega)\, ,
\label{cross_section}
\end{equation}
\end{widetext}
picking out the E1-E2 interference in the ${\bm p}\cdot{\bm A}$ interaction
between x rays and and electrons. 
The notation is as follows: $\hbar\omega$, ${\bf k}$ 
and ${\bm \epsilon}$ represent energy,  wave vector and 
polarisation of the photon; $\mid g\rangle$ and $\mid f\rangle$ 
denote ground and final states of the electron system, 
with energies $E_g$ and $E_f$ respectively; electrons are labelled 
by $\iota$ and $\iota'$; $\alpha = e^2/\hbar c$. 

The integrated intensity\cite{Note14}.
\begin{equation}
\int_{j_-+j_+}
\frac{\sigma^{\bm \epsilon}(\omega)}{(\hbar\omega)^2}
d(\hbar\omega)\,,
\label{master}
\end{equation}
expands into a linear combinations of pairs of irreducible tensors of 
increasing rank, $\mu=1,2,3$. 
Each pair is given by the scalar product between a wave-vector and 
polarisation response (geometrical factor) and the ground-state expectation 
value of an effective one-electron operator
\begin{displaymath}
\sum_q T^{(\mu,\pm)*}_q(\bm \epsilon, \hat{\bf k})\,
\langle g |{\cal O}^{(\mu,\pm)}_q(l,l')
| g \rangle\,. 
\end{displaymath}
In the second quantisation formalism, the latter is defined by
\begin{equation}
\sum_{\iota}{\cal O}^{(\mu,\pm)}_q (l,l')_{\iota}
=\sum_{mm'}\langle l'm' |
{\cal O}^{(\mu,\pm)}_q|lm \rangle 
a^{\dagger}_{l'm'}a_{lm}+{\rm c.c.}\, ,
\label{one_part}
\end{equation}
where $l'=l\pm1$ and $|lm\rangle$ stands for a spherical harmonic; 
$a^{\dagger}_{l m}$ and $a_{l' m'}$ create and annihilate valence electrons. 
All ${\cal O}^{(\mu,\pm)}(l,l')$ are odd under space inversion; their 
behaviour (even or odd) under time reversal is denoted by the superscript 
$\pm$.
 
Four operators ${\cal O}^{(\mu,\pm)}$ contribute to the E1-E2 absorption 
spectrum for arbitrary polarisation. They can all be written out in 
terms of three dimensionless order parameters: the familiar orbital 
angular momentum ${\bm L}$, an electric dipole ${\bm n}={\bf r}/r$, and 
${\bm \Omega} = ({\bm n}\times{\bm L} - {\bm L}\times{\bm n})/2$, a ME 
vector. 

By applying suitable recoupling transformations to 
Eq. (\ref{cross_section}), we obtain
\begin{widetext} 
\begin{equation}
{\cal O}^{(1,-)} = {\bm \Omega}\, ,\quad
{\cal O}^{(2,+)} =[{\bm L},{\bm \Omega}]^{(2)}\, ,
\quad
{\cal O}^{(2,-)} =[{\bm L},{\bm n}]^{(2)}\quad
{\rm and}\quad
{\cal O}^{(3,-)} = [[{\bm L},{\bm L}]^{(2)},{\bm \Omega}]^{(3)}\,,
\label{ord_par}
\end{equation}
\end{widetext}
where the couplings are defined via Clebsch-Gordan coefficients 
\begin{displaymath}
[U^{(p)}, V^{(\kappa)}]^{(k)}_q\equiv
\sum_{\mu,\nu}C^{kq}_{p\mu;\,\kappa\nu}U^{(p)}_{\mu}V^{\kappa}_{\nu}.
\end{displaymath}

Equations (\ref{ord_par}) identify one-electron properties revealed 
by E1-E2 absorption in the x-ray domain. The corresponding geometrical 
factors are given in Table \ref{TableII}. 
[As $T^{(1,+)}({\bm \epsilon},\hat{\bf k}) =
T^{(3,+)}({\bm \epsilon},\hat{\bf k})\equiv 0$, the polar moments  
${\cal O}^{(1,+)}={\bm n}$ and 
${\cal O}^{(3,+)}=[[{\bm L},{\bm L}]^{(2)},{\bm n}]^{(3)}$ 
do not contribute to expression (\ref{master}).]

Our derivation is based on a localised model, which considers a
single ion with a partially filled valence shell; all other shells
filled. Inclusion of spin-orbit interactions and/or crystal fields 
results in a deformation 
of the electronic cloud. Its multipolar expansion will contain spin 
and orbital moments characteristic of the symmetry of the deformation
and described by one-particle irreducible tensors. Valence states  
are hybridised in general, with the orbital part given as a  superposition
of two angular momenta: $l$ and $l'=l\pm1$, corresponding to $sp$ or 
$pd$ hybridisation in the actual case. (In our formalism, valence electrons 
are labelled using uncoupled orbital and spin quantum numbers.)
\begin{table}
\caption{\label{TableII} Polarisation responses of the E1-E2 interference.}
\begin{ruledtabular}
\begin{tabular}{l}
\\
$T^{(1,-)} (\hat{\bf k})_2= 
-\frac{1}{2}\sqrt{\frac{3}{5}}\hat{\bf k}$\\
\\
$T^{(2,+)} ({\bm \epsilon}, \hat{\bf k})_2
=\frac{\sqrt{3}}{2}[[{\bm \epsilon}^*,
{\bm \epsilon}]^{(1)},\hat{\bf k}]^{(2)}$\\
\\
$T^{(2,-)}({\bm \epsilon}, \hat{\bf k})_2=\frac{1}{2}
\left[\left[{\bm \epsilon}^*,
{\bm \epsilon}\right]^{(2)},\hat{\bf k}\right]^{(2)}$\\
\\ 
$T^{(3,-)} ({\bm \epsilon}, \hat{\bf k})_2
=\left[\left[{\bm \epsilon}^*,{\bm \epsilon}\right]^{(2)}
,\hat{\bf k}\right]^{(3)}$
\\
\\ \\
\end{tabular}
\end{ruledtabular}
\end{table}

As shown in previous work \cite{Tho92,Car93a,Car93b}, integrated dichroic 
spectra which stem from pure E1 or E2 transitions provide a measure 
of the ground-state expectation value of $I$-even irreducible tensors. 
Only the diagonal part (in the sense of $\delta_{l,l'}$) contributes to the 
ground-state matrix elements in this case.
Angular momentum  and parity are good quantum numbers; the known 
sum rules for linear and circular dichroism are recovered \cite{Note3}.  

Integrated E1-E2 dichroic spectra are related to the ground-state
expectation value of $I$-odd irreducible tensors. Only the off-diagonal
part contributes to the ground-state matrix elements, in this case. 
Orbital angular momentum is not a good quantum number and electron states 
do not have definite parity. Hybridisation effects are now 
observable and described by the operators ${\cal O}^{(\mu,\pm)}$.
 
Relations between XNCD, XNLD, XM$\chi$D and the order 
parameters given by Eqs. (\ref{ord_par}) will be provided below. 
\subsection{Interpretation of experiments}
\subsubsection{XNCD}
\begin{figure}
\includegraphics[width=8cm]{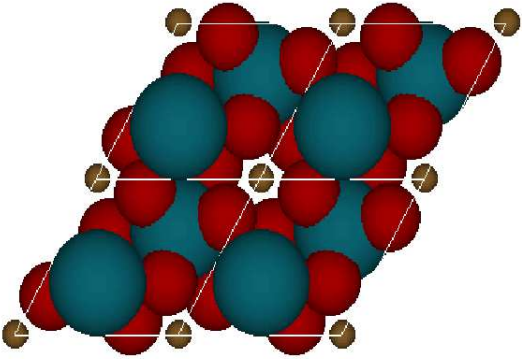}
\caption{\label{Fig.1} Crystal structure of $\alpha$-LiIO$_3$,
with \textcolor[rgb]{.5,.1,.1}{$\bullet$} = Li,
\textcolor[rgb]{.6,0,.05}{$\bullet$} = O, and 
\textcolor[rgb]{0,.5,.5}{$\bullet$} = I.}
\end{figure}

In this case, a recoupling of Eq. (\ref{cross_section}) leads to the 
following integral relation
\begin{widetext} 
\begin{equation}
\int_{j_-+j_+}\!\!\!\!\!\!\!\!\!\!\!\!
\frac{\sigma^{\bm \epsilon^+}_{\rm X}(\omega) - 
\sigma^{{\bm \epsilon}^-}_{\rm X}(\omega)}{(\hbar\omega)^2}
d(\hbar\omega)
=\frac{16\pi^2\alpha}{3\hbar c}(2l_c+1)
\sum_{l=l_c\pm1 \atop l'=l\pm1}\sum_{\iota,q }
R_{l_cl}^{(1)}R_{l_cl'}^{(2)}\,a^{(2,+)}_{l'}(l_c,l)
\sqrt{\frac{3}{2}}
T^{(2,+)*}_q({\bm \epsilon}^-,{\bm \epsilon}^+,\hat{\bf k})
\langle g | 
\left[{\bm L}_{\iota},{\bm \Omega}_{\iota}\right]^{(2)}_q(l,l')
| g \rangle \,  ,
\label{XNCD}
\end{equation}
\end{widetext}
which reveals the microscopic origin of XNCD spectra.
Here, ${\bm \epsilon}^{\pm}=\mp(i/\sqrt{2})
({\bm \epsilon}_1\pm i{\bm \epsilon}_2)$ define 
circular polarisation states; the index $\iota$ runs over valence
electrons \cite{Note4}. Furthermore,
\begin{widetext}
\begin{equation}
a^{(2,+)}_{l'}(l_c,l) 
=\frac{2(2l+1)(2l'+1)[6+3l_c(l_c+1)-2l(l+1)-l'(l'+1)]}{(l+l'+1)
(l_c-3l'+2l)(l_c+3l'-2l+1)
(l_c+l)^2(l_c+l+2)^2}
\label{a2+}
\, .
\end{equation}
\end{widetext}
The XNCD experiment of Goulon  {\it et al.} 
on $\alpha$-LiIO$_3$ \cite{Gou98}, whose crystal structure 
(space group $P6_3$) 
is depicted in Fig. \ref{Fig.1}, is readily interpreted with 
the help of Eq. (\ref{XNCD}).

We focus on charge properties stemming from $sp$
hybridisation. (The $4d$ shell of iodine is filled and the 
crystal displays no magnetism.) At the iodine sites the symmetry 
is $C_3$. Notice that ${\cal O}^{(2,+)} = 
\left[{\bm L},{\bm \Omega}\right]^{(2)}$, known as
{\it pseudodeviator} or {\it gyration} tensor, is totally 
symmetric in this point group {\it i.e.}, it is invariant
under all $C_3$ transformations. It yields therefore
a nonvanishing ground-state expectation value at each iodine site. 
The threefold axes of all the iodate groups are parallel in the unit
cell of $\alpha$-LiIO$_3$ ($P6_3$ is noncentrosymmetric). 
As a consequence, all the microscopic pseudodeviators add up 
upon absorption of circularly polarised x rays. This contribution
to the E1-E2 absorption profile is selected by circular dichroism
(see Table \ref{TableII}) yielding the observed XNCD spectrum.
The effect can thus be viewed as the $I$-odd analog of ferroquadrupolar
ordering, which is probed by pure E1 x-ray linear dichroism \cite{Car93b}. 
\subsubsection{XNLD}
In the case of XNLD, the integral relation reads
\begin{widetext}
\begin{eqnarray}
&&
\int_{j_-+j_+}
\!\!\!\!\!\frac{\sigma^{\parallel}_{\rm X}(\omega) - 
\sigma^{\perp}_{\rm X}(\omega)}
{(\hbar\omega)^2}
d(\hbar\omega)=\frac{8\pi^2\alpha}{\hbar c}
(2l_c+1)\sum_{l=l_c\pm1 \atop l'=l\pm1}
R_{l_cl}^{(1)}R_{l_cl'}^{(2)}
\sum_q\left\{\sqrt{\frac{2}{3}}
a^{(2,-)}_{l'}(l_c,l)\right. 
\nonumber
i\left( 
T^{(2,-)*}_q({\bm \epsilon}^{\parallel},\hat{\bf k})
-T^{(2,-)*}_q({\bm \epsilon}^{\perp},\hat{\bf k})\right)
\nonumber
\\
&& 
\langle g |\sum_{\iota} \left[
{\bm L}_{\iota},{\bm n}_{\iota}\right]^{(2)}_q(l,l')
| g \rangle
-2 a^{(3,-)}_{l'}(l_c,l)
\left( 
T^{(3,-)*}_q({\bm \epsilon}^{\parallel},\hat{\bf k})
-T^{(3,-)*}_q({\bm \epsilon}^{\perp},\hat{\bf k})\right)
\langle g |\sum_{\iota} \left[
[{\bm L}_{\iota},{\bm L}_{\iota}]^2,
{\bm \Omega}_{\iota}\right]^{(3)}_q(l,l')
| g \rangle
\left. 
\vphantom{\sqrt{\frac{2}{3}}}\right\}\, ,
\label{XNLD}
\end{eqnarray}
\end{widetext}
where 
$\parallel$ and $\perp$ denote two orthogonal linear-polarisation 
states. The expansion coefficients are given by 
\begin{equation}
a^{(3,-)}_{l'}(l_c,l) = \frac{a^{(2,+)}_{l'}(l_c,l)}
{6+3l_c(l_c+1)-2l(l+1)-l'(l'+1)}
\end{equation}
and 
\begin{equation}
a^{(2,-)}_{l'}(l_c,l)=\frac{(l+l'+1)(l'-l)}{2}a^{(2,+)}_{l'}(l_c,l)\, ,
\end{equation}
with $a^{(2,+)}_{l'}(l_c,l)$ defined by Eq. (\ref{a2+}).

Equation (\ref{XNLD}) reveals the microscopic ME 
nature of XNLD. (Notice the peculiarity of this form of linear
dichroism: {\it it changes sign upon reversal of an external 
applied magnetic field}\cite{Note17}. The result implies that Goulon and 
his co-workers have probed ME properties of (V$_{1-x}$Cr$_x$)$_2$O$_3$.  

\begin{figure}
\includegraphics[width=7cm]{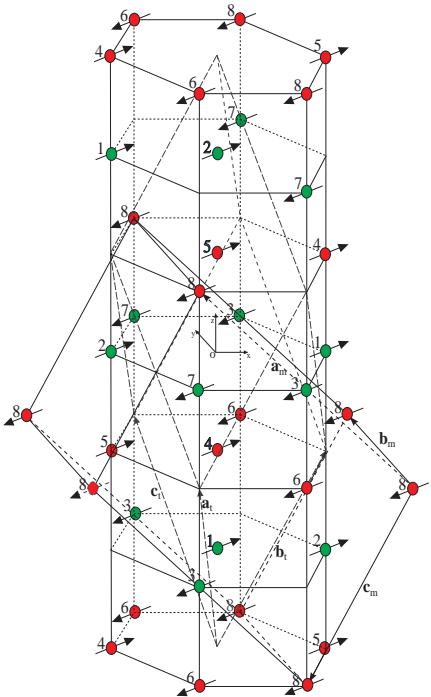}
\caption{\label{Fig.2} Corundum and monoclinic structures of V$_2$O$_3$.
(\textcolor[rgb]{1,0,0}{$\bullet$} and
\textcolor[rgb]{0,1,0}{$\bullet$} denote different orientations of
the oxygen octahedra.) The AFM structure is also
displayed.}
\end{figure}

V$_2$O$_3$ is characterised by a strongly destructive first-order
transition from a paramagnetic metallic phase (corundum structure)
to an antiferromagnetic (AFM) insulating phase (monoclinic) at 
T$_N\approx 150 ^{\circ}$K. The unit cells of pure V$_2$O$_3$ 
are depicted in Fig. \ref{Fig.2}.

\begin{figure}
\includegraphics[width=7cm]{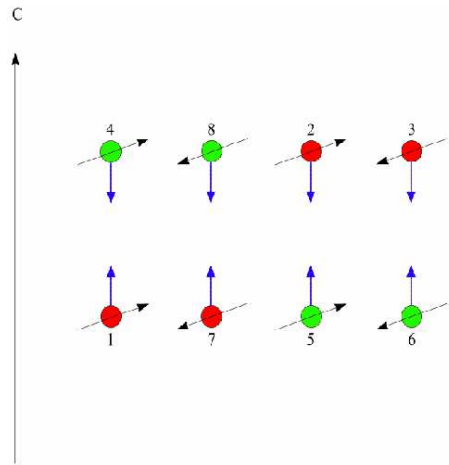}
\caption{\label{Fig.3} Antiferroelectric structure of 
(V$_{.972}$Cr$_{.028}$)$_2$O$_3$ at high temperature, as 
inferred from crystallographic data reported
in Ref. \cite{Der70}.}
\end{figure}

Substitution of Cr$^{3+}$ in the V$_2$O$_3$ lattice results in a 
Mott metal-to-insulator transition. At room temperature,  
charge properties of the compound can be deduced from structural studies,
which have been reported by Dernier and Marezio \cite{DaM70,Der70}. 
These authors ascribe the Cr-induced metal-to-insulator transition to an 
increase in the nearest neighbour vanadium-vanadium 
distances, which is accompanied by an 'umbrella-like' distortion of the 
oxygen octahedra. Crystal-structure refinements indicate that at each vanadium 
site the point group is C$_{3v}$ \cite{Der70}. Formation 
of an electric-dipole moment is thus permitted by symmetry. Crystallography 
also suggests that nearest-neighbour pairs of vanadium atoms 
are antiferroeletrically ordered along the hexagonal {\bf c} axis 
(the symmetry of a pair is D$_{3h}$ \cite{Der70}), 
as shown in Fig. \ref{Fig.3}. We can thus argue that, at about $300\, ^{\circ}$K 
in the insulating phase, (V$_{1-x}$Cr$_x$)$_2$O$_3$ is 
an antiferroelectric (AFE) crystal \cite{Note19}.  
\begin{figure}
\includegraphics[width=7cm]{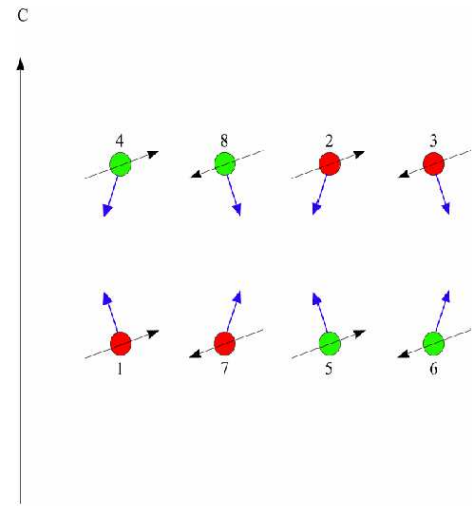}
\caption{\label{Fig.4} Tilting of the electric moments in the monoclinic
phase of (V$_{.972}$Cr$_{.028}$)$_2$O$_3$. (The arrows at about 71$^{\circ}$
with respect to the {\bf c} axis denote the magnetic moments.)}
\end{figure}

At $T = T_N$, the crystal makes a nearly second-order transition to an 
AFM monoclinic phase \cite{Mon70}. The driving mechanism of this 
transition is not fully understood \cite{Yet87}. The lattice space 
group lowers to $I2/a$, with magnetic point group $C_{2h}\otimes R$ 
\cite{DiM02}. Notice that this symmetry is incompatible
with the observation of XNLD. However, a reduction of magnetic 
symmetry leading to $C_{2h}(C_{1h})$ or $C_{2h}(C_2)$ might take 
place\cite{Note20}.

A lowering of magnetic symmetry could occur as follows.  
The structural change to the monoclinic phase eliminates the 
three-fold axis, which characterises the corundum structure. It
probably involves a co-operative Jahn-Teller distortion, which tilts the
electric moments. A tilting with the pattern depicted in Fig. \ref{Fig.4} 
would render (V$_{1-x}$Cr$_x$)$_2$O$_3$ an AFM, AFE, and ME crystal, thus
explaining the observations of Goulon and his collaborators. 

A physical interpretation of Eq. (\ref{XNLD}) is obtained by
expanding the "energy" function $W(\bm n,\bm L)$\cite{Bar82}. 
The $m$th term in the MacLaurin formula reads:
$
\frac{1}{m!}\left(n_{\alpha}\partial_{n_{\alpha}}
+L_{\beta}\partial_{L_{\beta}}\right)^m W(0,0)\, ;
$
here, greek letters denote cartesian components and repeated
indexes are summed over. It is readily seen that 
$\left[{\bm L},{\bm n}\right]^{(2)}$ stems from the $m=2$ 
term and describes a linear ME effect. The contribution
$\left[[{\bm L},{\bm L}]^2,
{\bm \Omega}\right]^{(3)}$ is found in the $m=4$ term and accounts
for a trilinear ME effect. 
\subsubsection{XM$\chi$D}
The third form of E1-E2 x-ray dichroism, namely XM$\chi$D,
was probed by Goulon and co-workers \cite{Gou02} 
at the cromium K-edge in Cr$_2$O$_3$ by 
implementing a new experimental technique:
{\it the dichroism of unpolarised x rays}. A microscopic
description of this spectroscopy, which is readily obtained from our
theory of integrated spectra, is reported in the current subsection.  

As is known \cite{Car93b}, the following integral relationship 
holds for pure E1 transitions 
\begin{widetext} 
\begin{equation}
\int_{j_++j_-} \frac{(\sigma^{{\bm \epsilon}^+}
+\sigma^{{\bm \epsilon}^-})_{E1}}{(\hbar\omega)^2}d(\hbar\omega)
=4\pi^2 \alpha \sum_l
\left[\frac{2}{3}\frac{n_h (l)}{2l+1} + b^{(2,+)}(l_c,l)
\langle g|\sum_{\iota}(3L^2_z-{\bm L}^2)_{\iota} |g\rangle
\right]
[P_{l_cl}^{(1)}]^2 
\, ,
\label{unpol} 
\end{equation}
\end{widetext}
where $n_h(l)= 4l+2- \langle g|n|g\rangle$ denotes the number 
of holes in the valence $l$ shell, and 
$P_{l_cl}^{(1)}=\langle c||\sqrt{4\pi/3}\,{\bf Y}^{(1)}||l\rangle
R^{(1)}_{l_c,l}$. Furthermore,
\begin{widetext}
\begin{equation}
b^{(2,+)}(l_c,l)=\frac{l(l+1)[3l(l+1)+1-6l_c(l_c+1)]+3l_c(l_c+1)
[l_c(l_c+1)-3]}{6l
(2l-1)(2l+1)(l+1)(2l+3)}
\end{equation}
\end{widetext}

In the case of a magnetoelectric crystal, E1-E2 corrections 
to Eq. (\ref{unpol}) need to be taken into account. They read
\begin{widetext}
\begin{eqnarray}
\int_{j_++j_-} \frac{(\sigma^{{\bm \epsilon}^+}
+\sigma^{{\bm \epsilon}^-})_{E1-E2}}{(\hbar\omega)^2}d(\hbar\omega)
&=& \frac{2\pi^2\alpha}{\hbar c}\,(2l_c+1)
\sum_{l=l_c\pm1 \atop l'=l\pm1}R_{l_cl}^{(1)}R_{l_cl'}^{(2)} 
\left[\frac{2}{5}\, a^{(1,-)}_{l'}(l_c,l)\, k_z\,
\langle g | \sum_{\iota}
{\bm \Omega}_{\iota}(l,l')_z | g \rangle \right.
\nonumber
\\
&-&
\left.
\frac{16}{\sqrt{10}}\, a^{(3,-)}_{l'}(l_c,l)\, k_z\, 
\langle g |\sum_{\iota} \left[
[{\bm L}_{\iota},{\bm L}_{\iota}]^2,
{\bm \Omega}_{\iota}\right]^{(3)}_z(l,l')
| g \rangle
\right]
\label{corrs}
\end{eqnarray}
\end{widetext}
with
\begin{equation}
a^{(1,-)}_{l'}(l_c,l) = \frac{(l_c+l+1)(l_c+l-2l')
h_1(l_c,l,l')}{(l_c+l)(l_c+l+2)(l+l'+1)^2}\, ,
\end{equation}
where $h_1(l_c,l,l') = l_c+2l'-l+1$ and $l' =l \pm 1$. We have set 
${\bf k} = k_z\, \hat{\bf z}$ in Eqs. (\ref{unpol} - \ref{corrs}).

Eqs. (\ref{unpol} - \ref{corrs}) indicate that XM$\chi$D is observed 
as follows. Record the $\sigma^{{\bm \epsilon}^+}
+\sigma^{{\bm \epsilon}^-}$ spectrum in a 
magnetoeletric cystal annealed in the parallel
configuration. Ditto for antiparallel annealing. Subtract the two
spectra. 

It goes without saying that such a dichroic spectrum, which is
described by Eq. (\ref{corrs}), reflects ME properties of crystals. 
\section{Discussion and outlook}
To provide a theoretical interpretation of recent x-ray experiments, 
which have detected XNCD, XNLD and XM$\chi$D respectively in $\alpha$-LiIO$_3$, 
(V$_{.972}$Cr$_{.028}$)$_2$O$_3$ and Cr$_2$O$_3$, 
the current paper has presented a theoretical analysis of  
near-edge absorption of polarised x rays in noncentrosymmetric 
crystals. Our work has centred on the derivation of integral
relations for dichroic effects which stem from E1-E2 inner-shell 
excitations. Such relations are written out in terms of one-electron 
effective operators (order parameters), which identify 
crystalline microscopic properties revealed by the observed spectra.
Two classes of parity-breaking order parameters have been found, which 
correspond to polar and magnetoelectric moments of valence electrons.
Only orbital degrees of freedom have been considered in our derivation; 
inclusion of spin would be straightforward. 

We remind the reader that two irreducible tensors, 
namely ${\cal O}^{(1,+)}={\bm n}$ and 
${\cal O}^{(3,+)}=[[{\bm L},{\bm L}]^{(2)},{\bm n}]^{(3)}$ 
(see Section \ref{ord-par}), have not entered our discussion
of integrated XNCD, XNLD and XM$\chi$D spectra. 
As previously observed, ${\cal O}^{(1,+)}$ and 
${\cal O}^{(3,+)}$ have vanishing geometrical factors and do not 
appear in our coupled expansion of expression (\ref{master}). They do
contribute however to x-ray resonant diffaction, as can be verified by
extending the work of Luo {\it et al.} \cite{Luo93} to include E1-E2
processes. (Such a lengthy derivation will not be reported here.)
The result is of interest as it indicates the possibiity of detecting 
{\it Bragg peaks from ferrolectric and AFE structures} with x rays at 
resonance. 

In conclusion, the pioneering experiments of Goulon and his collaborators
appear to have paved the way for new investigations of electronic states in
noncentrosymmetric crystals using x-ray absorption and resonant scattering. 
As shown by our work, a variety of one-electron properties, which result from
parity-nonconserving electron interactions, can be measured, yielding valuable
information in condensed matter physics and material science.
 
\begin{acknowledgments}
Stimulating discussions with F. de Bergevin, P. J. Brown, J. Goulon, 
and E. Katz are gratefully acknowledged. We would like to thank 
S. di Matteo for pointing out an error in an earlier version 
of Eq. (\ref{XNLD}). We are indebted to B. G. Searle for providing
the computer program that generated Fig. \ref{Fig.1}.
\end{acknowledgments}
\appendix*
\section{Symmetry Analysis}
Lie groups (or, alternatively, Lie algebras) furnish 
a powerful tool for 
interpreting x-ray dichroism in non-centrosymmetric crystals. 
In fact, effective microscopic operators, which express electronic
properties revealed by the spectra, are readily deduced 
from the pertinent group generators. 
Our approach hinges therefore on spectrum-generating algebras, 
a concept originally introduced in nuclear \cite{GoL59} 
and particle physics \cite{DGN65,DaN66}.

We follow Carra and Benoist\cite{CaB00} and consider the 
general framework of a de Sitter algebra: 
$so(3,2)$ \cite{GoL59,Eng72}.
A realisation of such an algebra is provided by the 
operators:
\begin{equation}
{\bm A}^-=i({\bm A}-{\bm A}^{\dagger})/2,\quad 
{\bm A}^+=({\bm A}+{\bm A}^{\dagger})/2,
\end{equation}
\begin{equation}
{\bm L} \quad {\rm and} \quad N_0,
\end{equation}
where ${\bm L}$ denotes the orbital angular momentum (in units 
of $\hbar$) and $N_0|lm\rangle=(l+\frac{1}{2})|lm\rangle$, with 
$|lm\rangle$ a spherical harmonic. Furthermore,  
\begin{equation}
{\bm A}={\bm n}\,f_1(N_0)+{\bm \nabla}_{\Omega}f_2(N_0)\, ,
\end{equation}
with ${\bm n}={\bf r}/r$, 
${\bm \nabla}_{\Omega}=-i{\bm n}\times{\bm L}$ and  
\begin{equation}
f_1(N_0)=(N_0 - 1/2)f_2(N_0)\, ,
\end{equation}
\begin{equation}
f_2(N_0)=\sqrt{(N_0-1)/N_0}.
\end{equation}
${\bm A}$ and ${\bm A}^{\dagger}$ 
are known as shift operators as their action on $|lm\rangle$ 
changes $l$ into $l\pm1$.

A physical interpretation of ${\bm A}^-$
is provided by the relation
\begin{widetext} 
\begin{equation}
{\bm \Omega} =
({\bm n}\times{\bm L}-{\bm L}\times{\bm n})/2
=i\left[{\bm n},{\bm L}^2\right]/2
=\frac{i}{2}\left({\bm \nabla}_{\Omega} - 
{\bm \nabla}_{\Omega}^{\dagger}
\right)
=\frac{1}{2}\frac{1}{\sqrt{N_0}}\left[ N_0, {\bm A}^- 
\right]_+\frac{1}{\sqrt{N_0}}\, ,
\label{anapole}
\end{equation}
\end{widetext}
where $[...]_+$ denotes an anticommutator. 
Eq. (\ref{anapole}) defines the (purely angular) 
{\it orbital anapole} \cite{Zel57,FaK80,BaB97}.

A physical representation of the generator ${\bm A}^+$ is provided
by
\begin{equation}
{\bm A}^+ = \sqrt{N_0} {\bm n} \sqrt{N_0}\, .
\label{elec_dip} 
\end{equation}

${\bm L}, {\bm n}$ and ${\bm \Omega}$ provide the building blocks of our
derivation. They form a triad of 
mutually orthogonal dimensionless vectors. Their nature (magnetic, 
electric and magnetoelectric, respectively) is readily inferred from 
behaviour under space inversion and time reversal.

Notice that ${\bm L}$ (rotations) and ${\bm \Omega}$ (boosts) 
generate a {\it homogeneous Lorentz group}: SO(3,1). 
Also ${\bm L}$ and $\sqrt{N_0}{\bm n}\sqrt{N_0}$, where $N_0 |lm\rangle = 
(l+\frac{1}{2})| lm \rangle$, generate SO(3,1). The homogeneous Lorentz 
group enters our derivation of $I$-odd operators in a natural way, given 
its deep interweaving with chirality. Notice that ${\bm L}$, ${\bm \Omega}$ 
and $\sqrt{N_0}{\bm n}\sqrt{N_0}$ provide a realisation of the 
$so(3,2)$ de Sitter algebra. This is the required 
symmetry extension, with respect to pure E1 or E2 transitions, that was 
mentioned in the Introduction.  
\end{document}